\pdfoutput=1
\documentclass{article}
\usepackage[pdftex]{graphicx}
\usepackage{arxiv}
\usepackage{subfigure}
\usepackage{color}
\usepackage{url}
\usepackage{bm}
\usepackage[psamsfonts]{amssymb}
\usepackage{algorithm}
\usepackage{comment}
\usepackage{amsmath}
\usepackage[noend]{algpseudocode}
\usepackage{wrapfig}
\usepackage{caption}
\usepackage{multirow}

\algnewcommand{\IIf}[1]{\State\algorithmicif\ #1\ \algorithmicthen \ \Return}
\algnewcommand{\EndIIf}{\unskip\ \algorithmicend\ \algorithmicif}
\newcommand{\ctext}[1]{\raise0.2ex\hbox{\textcircled{\scriptsize{#1}}}}

\title{Accelerating Distributed Deep Reinforcement Learning by
  In-Network Experience Sampling}

\author{
  Masaki Furukawa\\
  Keio University\\
  3-14-1 Hiyoshi, Kohoku-ku, Yokohama, Japan\\
  \texttt{furukawa@arc.ics.keio.ac.jp}\\
  \And
  Hiroki Matsutani \\
  Keio University\\
  3-14-1 Hiyoshi, Kohoku-ku, Yokohama, Japan\\
  \texttt{matutani@arc.ics.keio.ac.jp} \\
}

\begin{document}

\maketitle

\begin{abstract}
A computing cluster that interconnects multiple compute nodes is used
to accelerate distributed reinforcement learning based on DQN (Deep
Q-Network).
In distributed reinforcement learning, Actor nodes acquire experiences
by interacting with a given environment and a Learner node optimizes
their DQN model. 
Since data transfer between Actor and Learner nodes increases
depending on the number of Actor nodes and their experience size,
communication overhead between them is one of major performance
bottlenecks.
In this paper, their communication performance is optimized by using
DPDK (Data Plane Development Kit).
Specifically, DPDK-based low-latency experience replay
memory server is deployed between Actor and Learner nodes
interconnected with a 40GbE (40Gbit Ethernet) network.
Evaluation results show that, as a network optimization technique,
kernel bypassing by DPDK reduces network access latencies to a shared
memory server by 32.7\% to 58.9\%.
As another network optimization technique, an in-network experience
replay memory server between Actor and Learner nodes reduces access
latencies to the experience replay memory by 11.7\% to 28.1\% and
communication latencies for prioritized experience sampling by 21.9\%
to 29.1\%.
\end{abstract}

\section{Introduction}\label{sec:intro}
Reinforcement learning is a machine learning approach to acquire an
action policy that can maximize a long-term reward by repeating trial
and error in action and observation at a given environment.
Q-learning is a typical reinforcement learning method, where Q-value
means effectiveness of an action in a state.
By taking an action based on Q-value and observing the environment,
the Q-value is continuously updated in order to acquire an optimal
action policy.
DQN (Deep Q-Network) introduces a deep neural network called Q-network
to approximate the conventional Q-learning, and recently it has been
applied in various application domains, such as game AI and robot
control.
In this case, the reinforcement learning takes an action based on
Q-network, observes the environment, and updates the Q-network by deep
learning.
Since these steps are repeated until the Q-network training is
converged, it typically takes a time.
In this paper, we focus on a typical case of distributed reinforcement
learning, in which the first two steps (i.e., taking an action by
Q-network and observing the environment) are distributed over multiple
nodes in order to accelerate acquisition of the optimal action policy.

In distributed reinforcement learning systems using DQN \cite{apex}
\cite{gorila}, Actor process is in charge of the first two steps and
Learner process is in charge of the last step (i.e., updating
Q-network by deep learning).
State transitions experienced by multiple Actor processes are
accumulated in an experience buffer, called experience replay memory,
and Learner process samples these experiences from the memory in order
to update Q-network.
Since data transfer between Actor and Learner nodes increases
depending on the number of Actor processes, experience size, and
Q-network model size, their communication overhead is one of major
performance bottlenecks in such distributed reinforcement learning
systems.
To reduce the communication cost, network processing optimizations by
DPDK (Data Plane Development Kit) \cite{dpdk} are applied to a shared
memory server and experience replay memory server, both located
between Actor and Learner nodes.
In this paper, first, a distributed deep reinforcement learning system
inspired by \cite{apex} is implemented on a cluster of computers
interconnected with 40GbE (40Gbit Ethernet) and analyzed in terms of
network access overheads.
Then, DPDK-based low-latency shared memory server and experience
replay server are evaluated to demonstrate benefits of the proposed
network optimizations on distributed deep reinforcement learning.

This paper is organized as follows.
Section \ref{sec:related} gives background on distributed deep
reinforcement learning, prioritized experience replay, and DPDK-based
network optimization techniques.
A distributed deep reinforcement learning system is implemented and
analyzed in terms of network overheads in Section \ref{sec:design},
and the proposed network optimizations are applied to the system in
Section \ref{sec:impl}.
Section \ref{sec:eval} evaluates the proposed system and Section
\ref{sec:conc} concludes this paper.

\section{Background}\label{sec:related}

\subsection{Distributed Deep Reinforcement Learning}
\begin{figure*}[h]
  \begin{center}
    \includegraphics[height=35mm]{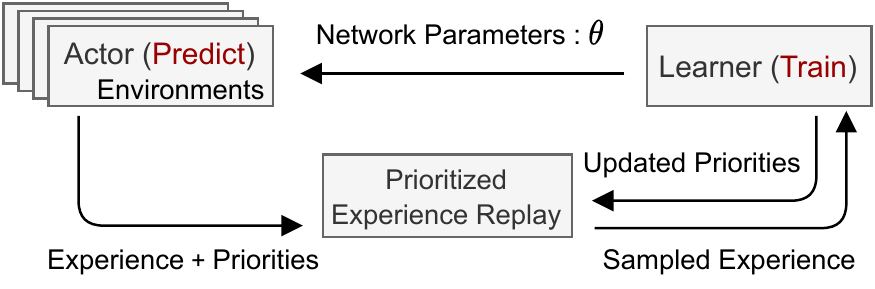}
    \caption{Ape-X architecture}
    \label{fig:apex}
  \end{center}
\end{figure*}

High-performance distributed deep reinforcement learning systems have
been widely studied recently.
Distributed Prioritized Experience Replay (Ape-X) \cite{apex} is one
of major architecture among them, and it is used as a baseline
distributed deep reinforcement learning architecture in this paper.
Figure \ref{fig:apex} illustrates Ape-X architecture.
Ape-X introduces a prioritized experience replay for large-scale
distributed reinforcement learning systems that consist of Actor
processes, experience replay memory, and Learner process.
Actor processes select actions based on Q-network inferences and
generate state transitions by the selected actions.
The state transitions or experiences are stored in an experience
replay memory, and the Q-network model is updated based on sampled
experiences by Learner process.
Roles of these processes are explained in the following subsections.

\subsubsection{Actor}
Actor is a process that takes actions based on DQN inferences and
observes rewards from environment to generate experiences each of
which consists of the original state, action, reward, and next state.
Algorithm \ref{algo:actor} shows Actor's behavior.
Actor makes inferences using a model parameter $\theta$ obtained from
Learner to select an action $a_t$ at current time $t$ (\ctext{1}).
$\varepsilon$-greedy is well-known approach to select an action $a_t$
from a set of possible actions $A$.
It can add diversity to the action search by randomly selecting an
action with a probability of $\varepsilon$.
Different $\varepsilon$ value is set to each Actor.
Actor then takes the selected action $a_t$ and observes reward $r_t$
and next state $s_{t+1}$ from the environment so that an experience
$(a_t, s_t, r_t, s_{t+1})$ is generated (\ctext{2}).
The generated experiences are temporarily stored in a local buffer
(\ctext{3}), and then those of a predefined batch size are transferred
to an experience replay memory (\ctext{5}).
A priority is assigned to each experience at Actor so that experiences
that can accelerate the DQN training are preferentially trained by
Learner (\ctext{4}).
A difference in Q-function between successive steps, called TD error,
is used as the priority of experience.
In DQN, a priority $p_t$ of an experience is calculated based on the
TD error $\delta_t$ as follows.
\begin{equation}
  p_t = |\delta_t| = |Q(s_t, a_t; \theta) - Q(s_{t-1}, a_{t-1}; \theta)|
\end{equation}

\begin{algorithm}[h]
  \caption{Actor}
  \label{algo:actor}
    \begin{algorithmic}
    \State \textbf{Pull} Parameters $\theta_0$ \hspace{\fill} $\blacktriangleright$ Get latest network parameters
    \For {$t = 1$ to $T$}
    \State $a_t \leftarrow \varepsilon - $greedy$(A)$ \hspace{\fill} {$\cdots \ \ctext{1}$}
    \State $(r_t, s_{t+1}) \leftarrow$ Environment$(a_t, s_t)$ \hspace{\fill} {$\cdots \ \ctext{2}$}
    \State LocalBuffer.ADD$((s_t, a_t, r_t, s_{t+1}))$ \hspace{\fill} {$\cdots \ \ctext{3}$}
    \If{LocalBuffer.SIZE() $\geq$ Batch\_Size}
      \State $\tau \leftarrow$ LocalBuffer.GET(Batch\_Size)
      \State $p \leftarrow$ ComputePriorities($\tau$) \hspace{\fill} {$\cdots \ \ctext{4}$}
      \State \textbf{Push}$(\tau, p)$ \hspace{\fill} {$\cdots \ \ctext{5}$}
    \EndIf
    \State \textbf{Pull} $\theta_t$ every $N_{pull}$ steps\hspace{\fill} $\blacktriangleright$ Get latest parameters {$\cdots \ \ctext{6}$}
    \EndFor
\end{algorithmic}
\end{algorithm}

\subsubsection{Learner}
Algorithm \ref{algo:learner} shows Learner's behavior.
Learner samples the experiences accumulated in the experience replay
memory based on their priorities assigned by Actor (\ctext{7}).
Learner uses the sampled experiences of a training batch size as
training data, and then it updates parameter $\theta$ of Q-function so
that a loss function of the training data is minimized (\ctext{8}).
More specifically, in DQN training, parameter of Q-function is updated
so that it can predict a sum of the latest reward $r_{t+1}$ and the
maximum expected reward in the next state $s_{t+1}$ as follows.
\begin{equation}
  Q(s_t, a_t) \leftarrow r_{t+1} + \gamma \max_{a_{t+1} \in A} A(a_{t+1}, s_{t+1}),
\end{equation}
where $\gamma$ is a discount rate of reward.
Since priorities of experiences once used in training should be
decreased, Learner updates priorities of such experiences in the
experience replay memory (\ctext{9}).
Learner sends the updated model parameter $\theta$ to Actor
(\ctext{10}), and Actor periodically updates the model with the latest
parameter (\ctext{6}).

\begin{algorithm}[h]
  \caption{Learner}
    \begin{algorithmic}
    \State $\theta_0 \leftarrow$ Initialized Parameters \hspace{\fill} $\blacktriangleright$ Set network parameters
    \For {$t = 1$ to $T$}
    \State $id, \tau \leftarrow$ SAMPLING(Batch\_Size) \hspace{\fill} {$\cdots \ \ctext{7}$}
    \State $l_t \leftarrow$ ComputeLoss$(\tau; \theta_{t-1})$ 
    \State $\theta_t \leftarrow$ UpdateParameters$(l_t;\theta_{t-1})$ \hspace{\fill} {$\cdots \ \ctext{8}$}
    \State UpdatePriorities$(id)$ \hspace{\fill} {$\cdots \ \ctext{9}$}
    \State \textbf{Set} $\theta_t$  \hspace{\fill} {$\cdots \ \ctext{10}$}
    \State Update $\theta$ every $N_{update}$ steps \hspace{\fill} $\blacktriangleright$ Fixed target Q-network
    \EndFor
  \end{algorithmic}
  \label{algo:learner}
\end{algorithm}

\subsubsection{Prioritized Experience Replay}
Although it is expected that to use experiences or state transitions
with higher priorities preferentially can improve efficiency of the
training, there are some issues.
That is, experiences with lower priorities may not be used for a long
time and an overfitting which becomes sensitive to noises may occur
due to a limited diversity of trained experiences.
To address these issues, a probabilistic sampling of experiences based
on their priorities \cite{prio} is used in recent distributed deep
reinforcement learning.
A sampling probability of an experience (or state transition) $i$ is
calculated based on priority $p$ of the experience as follows.
\begin{equation}
  P_{i} = \frac{p_i^\alpha}{\sum_k {p_k^\alpha}} \qquad (p_k \neq 0),
\end{equation}
where $\alpha$ is a hyper-parameter that weights the priority.

\begin{figure*}[h]
  \begin{center}
    \includegraphics[height=35mm]{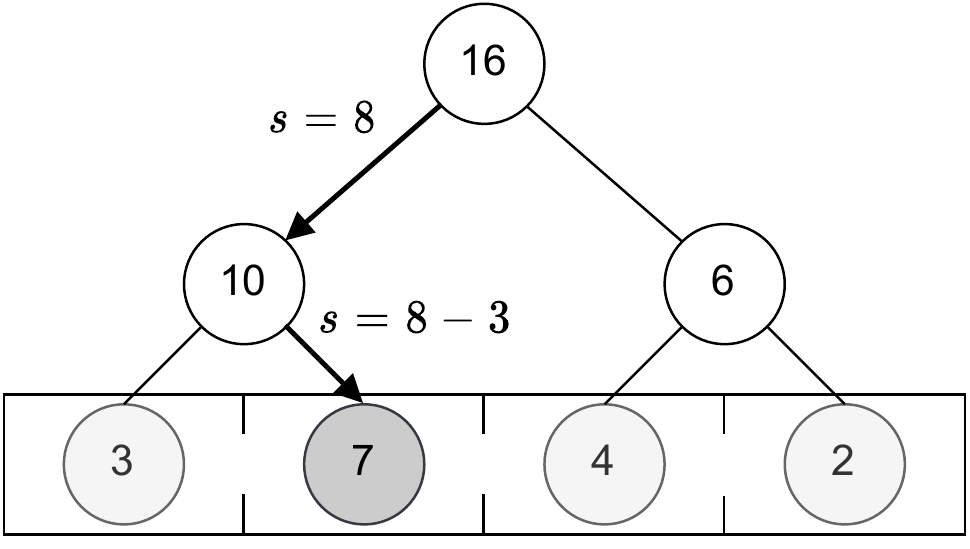}
    \caption{Example of experience sampling using SumTree}
   \label{fig:tree}
  \end{center}
\end{figure*}

In the prioritized experience replay, data manipulation and
probabilistic sampling of experiences can be efficiently implemented
by using SumTree as a data structure.
Algorithm \ref{algo:replay} shows the probabilistic sampling using
SumTree.
Figure \ref{fig:tree} illustrates an example of the experience
sampling using a random number $s=8$ for four priorities stored in
leaf nodes of SumTree.
By traversing the tree structure from root to leaf as described in
Algorithm \ref{algo:replay}, a probabilistic sampling based on
priority is implemented without reordering the experiences.
Computational complexity of the probabilistic sampling is $O(\log N)$.

\begin{algorithm}[h]
  \caption{Probabilistic experience sampling using SumTree}
    \begin{algorithmic}
    \Require{$0 \leq s$ (random number) $\leq \sum_{k}{p_{k}}$}
    \Require{$n : root$}
      \Function{Sampling}{$n, s$}
        \IIf{$n$ is leaf\_node} $n$
        \If{$n$.left.val $\geq s $} 
          \State \Return \Call{Sampling}{$n$.left, $s$}
        \Else 
          \State \Return \Call{Sampling}{$n$.right, $s-n$.left.val}
        \EndIf
    \EndFunction
  \end{algorithmic}
  \label{algo:replay}
\end{algorithm}

\subsection{Network Optimization using DPDK}\label{related:fstack}
\begin{figure}[h]
  \begin{center}
    \includegraphics[width=55mm]{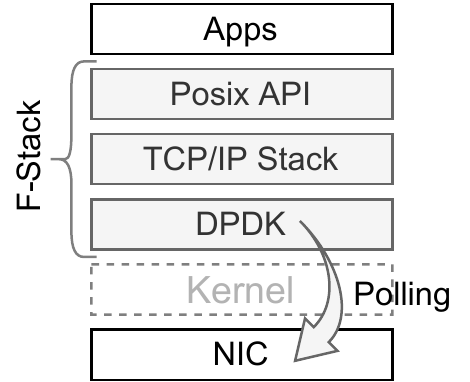}
    \caption{Kernel bypassing by DPDK and light-weight user-space
      TCP/IP stack by F-Stack}
    \label{fstack}
  \end{center}
\end{figure}

DPDK (Data Plane Development Kit) \cite{dpdk} is a well-known
framework for accelerating network processing by bypassing a network
protocol stack of OS kernel, as shown in Figure \ref{fstack}.
It dedicates specific CPU cores to network processing so that
user-space applications can directly access NIC (Network Interface
Card) in a polling manner.
By the polling based access to NIC without intervention of OS kernel,
network processing overheads due to frequent context switching and
data copy needed for packet send/receive events can be mitigated for
enabling a low-latency and high-throughput network processing.
F-Stack \cite{fstack} is a high-performance network processing
framework in cooperation with DPDK, which offers an embedded TCP/IP
stack borrowed from FreeBSD and co-routine APIs to user-space DPDK
applications, as shown in Figure \ref{fstack}.
DPDK-ANS \cite{ans} is also known as a similar framework.

In this paper, a shared memory parameter server and a prioritized
experience replay memory server of distributed deep reinforcement
learning systems are accelerated by using F-Stack and DPDK.
As a similar work, a DPDK-based network optimization is applied to a
distributed deep learning system \cite{my}, in which a network switch
is in charge of gradient aggregation of distributed deep learning and
its network processing is accelerated by DPDK.
In addition to the gradient aggregation, parameter optimization
algorithms, such as SGD, Adagrad, Adam, and SMORMS3, are accelerated
by in-network FPGA switch in \cite{itsubo}.
Please note that our target distributed deep reinforcement learning
employs a different architecture compared to typical distributed deep
learning frameworks such as those using AllReduce algorithm \cite{horovod}.
Specifically, experiences are exchanged in our system, while typical
distributed deep learning frameworks transfer gradients, which are
more likely to be outdated; thus a different approach is needed for
optimizing our target distributed deep reinforcement learning system.


\section{Baseline Distributed Deep Reinforcement Learning System}
\label{sec:design}

\subsection{Design and Implementation}
\begin{figure*}[h]
  \begin{center}
    \includegraphics[width=120mm]{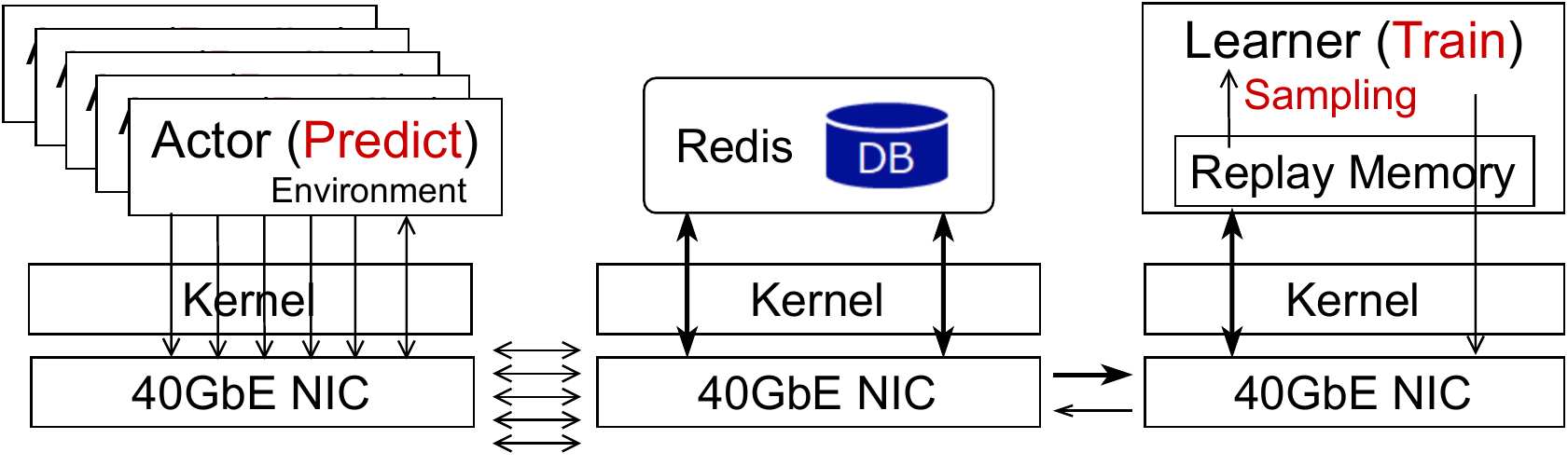}
    \caption{Baseline distributed deep reinforcement learning system
      in which Actor, shared memory (Redis), and Learner are
      implemented on different machines which are interconnected with
      40GbE switch}
    \label{fig:system}
  \end{center}
\end{figure*}

Figure \ref{fig:system} illustrates a baseline distributed deep
reinforcement learning system used in this paper.
From the left, Actor node, shared memory server, and Learner node are
illustrated in this figure.
In distributed deep reinforcement learning systems, functionalities to
share experiences and Q-network model between Actor and Learner nodes
are typically needed.
In our baseline system, the shared memory server is implemented with
Redis \cite{redis} which is a fast in-memory database system.
Actor nodes, shared memory server, and Learner node are running on
different machines which are interconnected with a fast 40GbE network.
The number of Actor nodes and the number of Actor processes running on
a single Actor node can be increased.
As for the Learner, a training process that updates Q-network
parameter and its child process that replays prioritized experiences
are running on the single Learner node.
Multiple Actor processes push their experiences and priorities of a
predefined batch size to Redis server, and the experience replay
memory server pulls them periodically.
The training process of Learner updates a deep neural network model by
using batch of experiences sampled from the experience replay memory
and then the updated model parameter is set to Redis.
Actor and Learner are implemented with Python language and PyTorch as
a machine learning library.
One and two GPUs are used at Actor and Learner nodes, respectively, to
accelerate the deep neural network inference and training.
Redis server is implemented with C language. 
Actor and Learner access Redis server by using Python interface of
Redis.
The prioritized experience replay memory is implemented with SumTree
data structure.
Machine specification of the baseline implementation is listed in
Table \ref{tab:env} of Section \ref{sec:eval}.

\subsection{Preliminary Evaluations}
\begin{figure*}[h]
  \begin{center}
    \includegraphics[width=130mm]{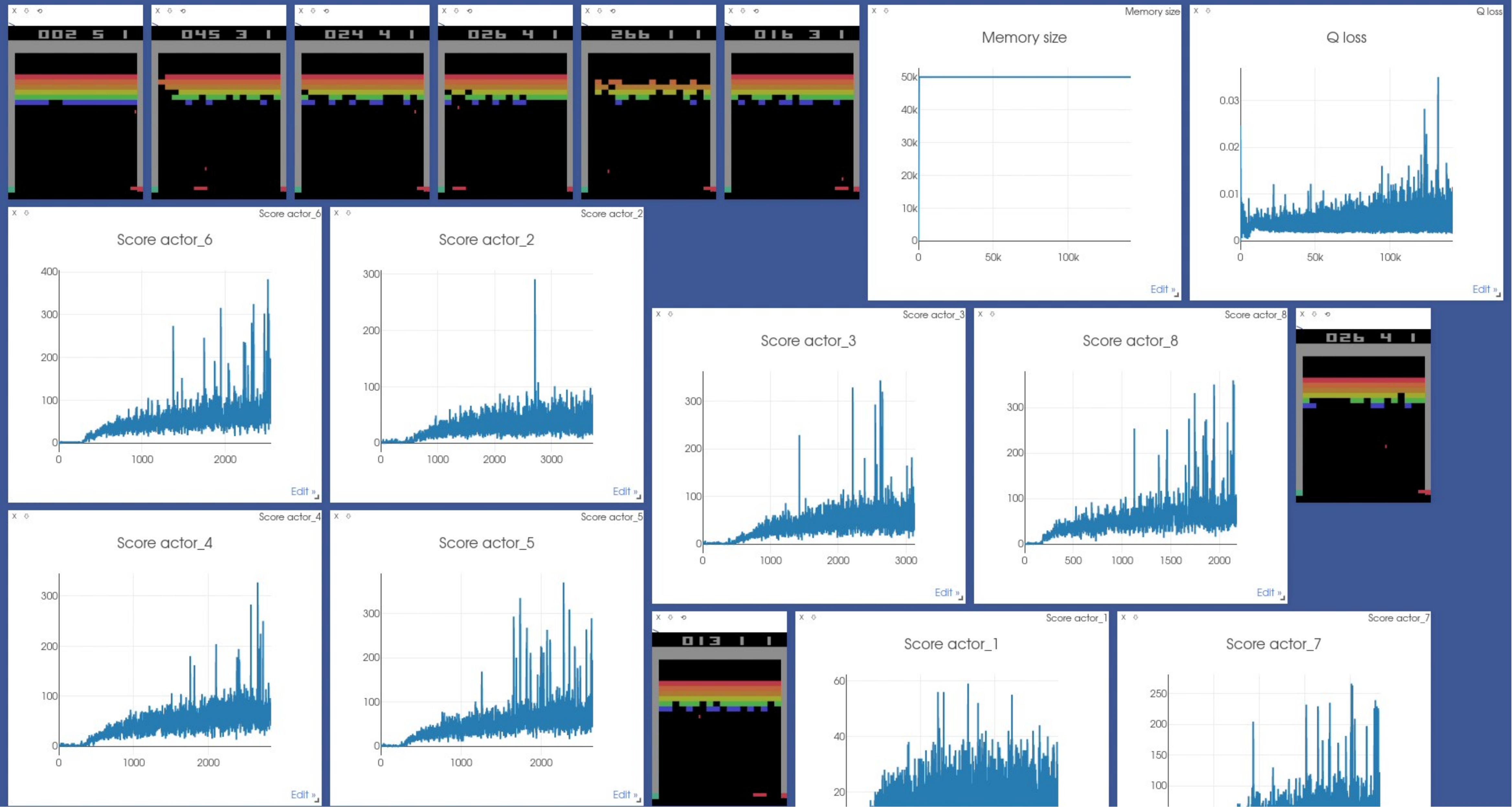}
    \caption{Screenshot of distributed deep reinforcement learning
      system when Breakout game is trained by eight Actors}
    \label{fig:break}
  \end{center}
\end{figure*}

The baseline distributed deep reinforcement learning system is
analyzed to see network access overheads.
OpenAIGym \cite{gym} is a well-known reinforcement learning simulation
environment, and Atari's Breakout game in the environment is used as
a reinforcement learning task in this analysis.
Figure \ref{fig:break} shows a screenshot of our distributed deep
reinforcement learning system when Breakout game is trained by using
eight Actor processes.
A game screen (background color: black) and score graph (background
color: white) are displayed for each Actor in our
management console implemented with Visdom \cite{visdom}.
Dueling Network Architecture \cite{duelling} is a well-known deep
neural network model used in DQN such as in \cite{apex}.
It is used in our baseline in cooperation with double-DQN and
$n$-step bootstrap target ($n=3$) techniques.
Input images (game screens in Figure \ref{fig:break}) are gray-scaled
and resized to $84 \times 84$, and four
resized frames are combined as a single input data; thus, the number
of input nodes of the network is $4 \times 84 \times 84$.
The number of actions is four, and thus, the number of output nodes is
four.

\begin{figure*}[h]
  \begin{center}
    \includegraphics[width=140mm]{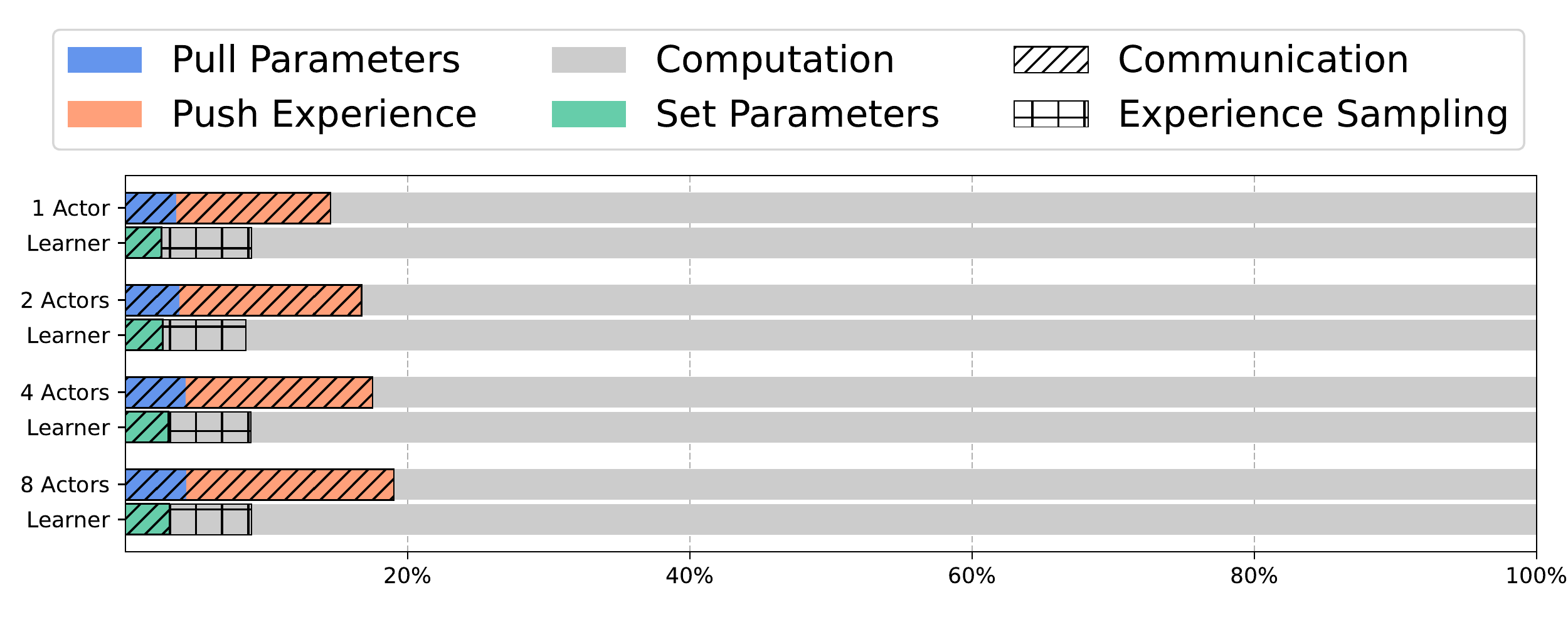}
    \caption{Execution time breakdown of baseline distributed deep
      reinforcement learning system}
    \label{fig:bar1}
  \end{center}
\end{figure*}

Figure \ref{fig:bar1} shows execution time breakdown when the number
of Actor processes is changed from one to eight.
Each result consists of two bar graphs: Actors' breakdown (upper) and
Learner's breakdown (lower).
The Actors' breakdown includes computation time, push experiences
time, and pull parameters time.
The Learner's breakdown includes computation time, experience sampling
time, and set parameters time.
Batch size of experiences that Actor processes send at a time is set
to 200 (approximately 42.7MB).
Parameter size of a deep neural network model is approximately 13MB.
Pull frequency of the parameters is once in 200 steps.
Training batch size of experiences at Learner is set to 512, and the
experience replay memory size is 65,536.
These values are tuning parameters and their optimal values vary
depending on a given environment.

In the graph, the more computation time the better.
Although communication time for model parameters (i.e., pull parameters
and set parameters) is not dominant, that for experiences
at Actor nodes (i.e., push experiences) increases as the number of
Actor processes is increased.
Even when machines are interconnected by a high-bandwidth 40GbE
network, network access latencies degrade performance of a distributed
deep reinforcement learning system.
To reduce communication overheads in distributed deep reinforcement
learning systems, a shared memory server and experience replay memory
server where communication is concentrated are accelerated in the next
section.


\section{Network Optimization on Distributed Deep Reinforcement Learning}
\label{sec:impl}

\subsection{Low-Latency Shared Memory by DPDK}\label{impl:fredis}
\begin{figure*}[h]
  \begin{center}
    \includegraphics[width=120mm]{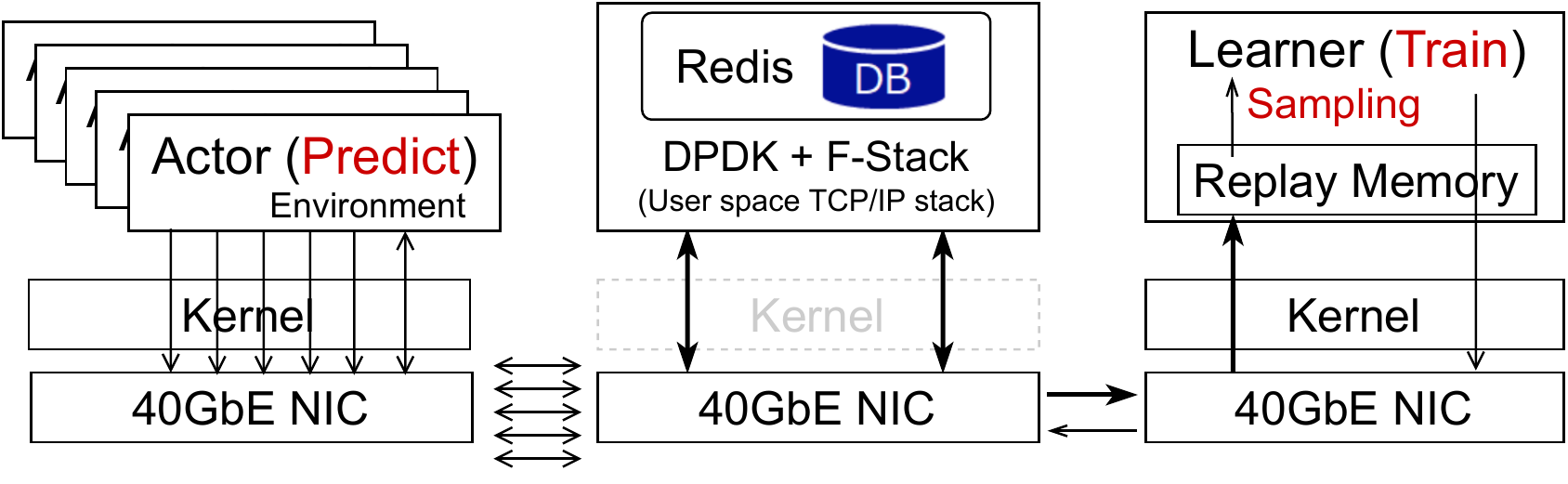}
    \caption{Distributed deep reinforcement learning system where DPDK
    is applied to shared memory server}
    \label{fig:proposal1}
  \end{center}
\end{figure*}

As the first network optimization, network access latency of a shared
memory is reduced by applying DPDK to Redis server.
In DPDK, dedicated threads running on specific CPU cores are polling
the NIC, and user-space applications can directly access the NIC
without intervention of OS kernel; thus, network processing overheads
are reduced.
Figure \ref{fig:proposal1} illustrates an optimized distributed deep
reinforcement learning system where DPDK is applied to the shared
memory server (Redis).
In this implementation, a low-latency shared memory server is built
with DPDK, F-Stack, and F-Stack compatible Redis.
In the center of figure, packets received by the NIC of shared memory
server bypass TCP/IP stack of OS kernel so that they directly go to
Redis server running on the application layer.
Instead, a light-weight TCP/IP stack of F-Stack is used to access
Redis server.
It is used as a user-space network protocol stack in cooperation with
DPDK as mentioned in Section \ref{related:fstack}.
In this figure, the central node is exclusively used for F-Stack
compatible Redis server, in which NIC port and CPU cores are occupied
by DPDK and F-Stack processes.
As F-Stack compatible Redis server, an implementation of \cite{fredis}
is used in this paper.
Please note that only the Redis server is changed in this
implementation.
Actors, Learner, and experience replay memory are not modified from
the baseline distributed deep reinforcement learning system.

\subsection{Low-Latency Experience Replay Memory by DPDK}
\begin{figure*}[h]
  \begin{center}
    \includegraphics[width=120mm]{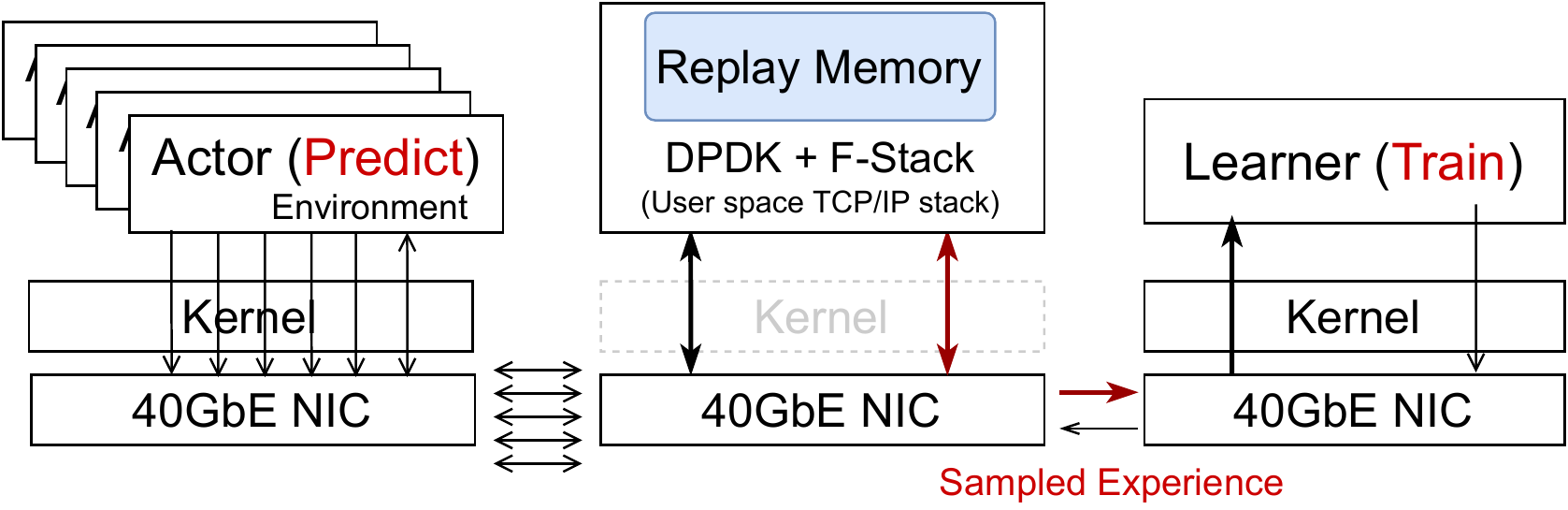}
    \caption{Distributed deep reinforcement learning system where
      experience replay memory is co-located with low-latency shared
      memory server}
    \label{fig:proposal2}
  \end{center}
\end{figure*}

By introducing a low-latency shared memory server by DPDK, it is
expected that experiences transfer throughput of Actors (i.e., push
frequency of experiences) is improved.
However, as Actor's throughput is improved (e.g., using GPUs) and the
number of Actors is increased, the number of experiences accumulated
in a shared memory per time is also increased.
Since an experience replay memory periodically pulls all the
experiences from the shared memory, data transfer time by the
experience replay memory would be a performance bottleneck when the
number of experiences in the shared memory is increased.
To eliminate this bottleneck, as the second network optimization, an
experience replay memory is implemented in the same machine of the
low-latency shared memory server by DPDK.
Figure \ref{fig:proposal2} illustrates another optimized distributed
deep reinforcement learning system where an experience replay memory
is co-located with the low-latency shared memory server.
Experiences sent by multiple Actors are sampled at the low-latency
shared memory server (i.e., central node) implemented with DPDK.
In this case, already sampled experiences of the training batch size
are only transferred to the Learner node, and thus data transfer
between the shared memory node and Learner node can be significantly
reduced.

Since the F-Stack compatible Redis server is directly bind to NIC port
without intervention of OS kernel, only packets sent to that NIC port
from network can access the Redis server, which means that an
experience replay memory process implemented on the same machine
cannot access the F-Stack compatible Redis.
Because of this limitation, in the second optimized implementation, we
do not use the F-Stack compatible Redis as a low-latency shared memory
server.
Instead, we newly implement an F-Stack compatible experience replay
memory that includes in-house shared memory functionality similar to
Redis.
In this case, since APIs provided by F-Stack are implemented with
C/C++ language, our experience replay memory is also implemented with
C/C++ language.
In addition, Actor and Learner are modified to use ctypes structure in
their communication protocol so that they can communicate with the new
experience replay memory server implemented with C/C++ language.

\begin{figure*}[h]
  \begin{center}
    \includegraphics[width=120mm]{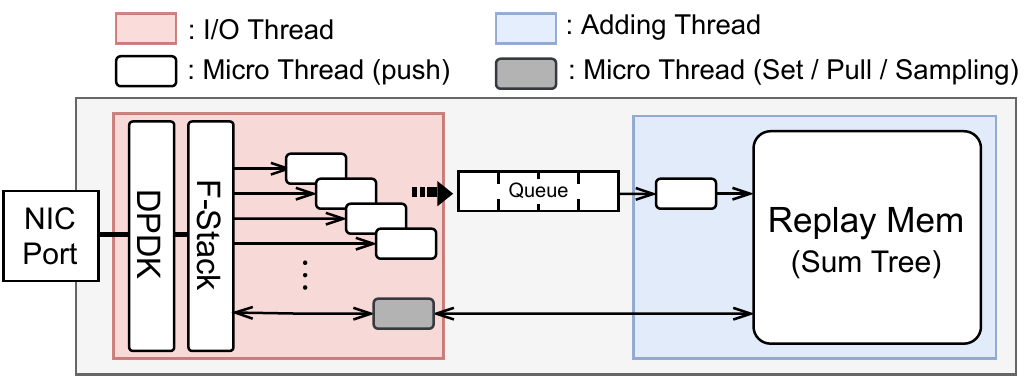}
    \caption{Experience replay memory server including shared memory
      implemented with DPDK and F-Stack}
    \label{fig:impl}
  \end{center}
\end{figure*}

Figure \ref{fig:impl} illustrates new experience replay memory server
including a shared memory implemented with DPDK and F-Stack.
A CPU core dedicated to F-Stack is polling a specific NIC port in
order to accelerate the network processing.
As shown in Figure \ref{fig:proposal2}, multiple Actor processes are
running on a single Actor node (i.e., left node).
In the central node, a micro-thread is launched for each Actor process
to multiplex the network I/O at the experience replay memory server as
detailed in Figure \ref{fig:impl}.
These micro-threads are implemented by using co-routine APIs
\cite{micro} provided by F-Stack.

By using SumTree as a data structure of experience replay memory,
computational complexity of a prioritized probabilistic sampling is
$O(\log N)$ when the number of experiences is $N$.
Please note that adding an experience to SumTree takes a computation
cost of $O(\log N)$ and occupies a CPU core, which may prevent the
polling-based network processing.
To address this issue, a separate thread dedicated for adding
experiences to SumTree (i.e., adding thread) is implemented as shown
in Figure \ref{fig:impl}.
I/O micro-threads and adding thread communicate via a queue structure
in the experience replay server.
This queue structure in this second optimized implementation is
corresponding to Redis server in the first optimized implementation of
Figure \ref{fig:proposal1}.


\section{Performance Evaluations}\label{sec:eval}
The following two network optimizations on a distributed deep
reinforcement learning system are evaluated in terms of network
processing latencies.
\begin{enumerate}
\item Low-latency shared memory by DPDK (Figure \ref{fig:proposal1})
\item Low-latency experience replay memory including shared memory by
  DPDK (Figure \ref{fig:proposal2})
\end{enumerate}
Here, an ideal evaluation metric is an execution time required to
complete the Breakout game by our reinforcement learning system.
According to \cite{apex}, five to six hours are typically required to
get a high score of the Breakout game when the number of Actor
processes is 360; however, only up to ten Actor processes can be
executed in our environment due to machine limitation, which means
that the same evaluation metric is not feasible in this paper.
Instead, the distributed deep reinforcement learning system before and
after the network optimizations is evaluated in terms of 1) access
latency to the shared memory, and 2) access latency to the experience
replay memory.
Table \ref{tab:env} shows machines used in the evaluation environment.

\begin{table*}[h]
\begin{footnotesize}
  \centering
  \caption{Evaluation environment}
  \begin{tabular}{cccc}
    \hline 
     & Actors machine & Shared/Replay memory machine & Learner machine \\
    \hline
    OS       & Ubuntu 20.04 & Ubuntu 20.04 & Ubuntu 20.04 \\
    CPU      & Intel Xeon E5-2637 v3 @3.5GHz & Intel Xeon CPU E5-2637 v3 @3.50GHz & Intel Xeon CPU E5-1620 v2 @3.50GHz \\
    Memory   & 128 GB & 512 GB & 128 GB\\
    GPU      & GeForce RTX 3080 $\times 1$ & $-$ & GeForce GTX 1080Ti $\times 2$ \\
    CUDA/PyTorch & 11.3 / 1.8.0+cu111 & 11.3 / 1.8.0+cu111 & 11.3 / 1.8.0+cu111 \\
    NIC	     & Intel Ethernet CNA XL710-QDA2 & Intel Ethernet CNA XL710-QDA2 & Intel Ethernet CNA XL710-QDA2\\
    DPDK     & $-$ & 20.11.0 & $-$ \\
    Redis    & $-$ & 5.0.5 & $-$ \\
    Network  & \multicolumn{3}{c}{Mellanox 40G Switch SX1012} \\
    \hline
  \end{tabular}
  \label{tab:env}
\end{footnotesize}
\end{table*}

\subsection{Low-Latency Shared Memory by DPDK}\label{eval:fredis}
\begin{figure*}[h]
  \begin{center}
    \includegraphics[height=78mm]{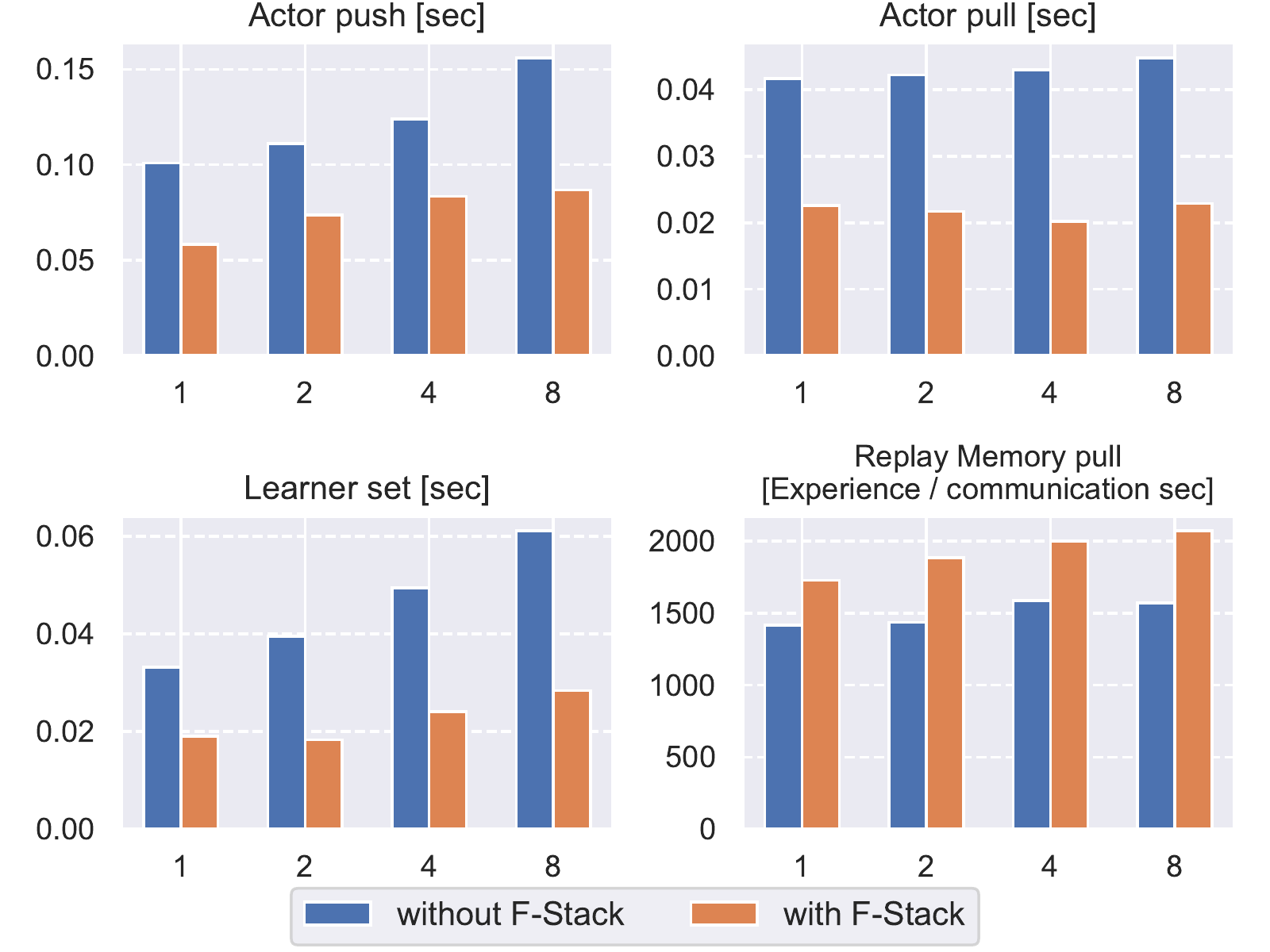}
    \caption{Evaluation result of access latency to shared memory
      (first optimized implementation)}
    \label{fig:result1}
  \end{center}
\end{figure*}

Figure \ref{fig:result1} shows Redis access latencies of Actor,
Learner, and experience replay memory in the first optimized
implementation (see Figure \ref{fig:proposal1}) when the number of
Actor processes is changed from one to eight.
The upper left, upper right, and lower left graphs show the push
latency of Actor, pull latency of Actor, and set latency of Learner,
respectively.
X-axis is the number of Actor processes and Y-axis is elapsed time in
second.
The lower right graph shows the number of experiences pulled by the
experience replay memory per a communication time.
In other words, it is the average number of experiences pulled from
the shared memory during a pull communication of one second.
By introducing the F-Stack based low-latency shared memory, the push
latency of Actor, pull latency of Actor, and set latency of Learner
are reduced by 32.7\%-44.4\%, 45.8\%-58.0\%, and 42.8\%-58.9\%,
respectively.
Since access latencies to the shared memory are reduced, the number of
experiences pulled by the experience replay memory per a communication
time is increased by 21.9\%-31.9\%.
Although access latencies typically increase as the number of Actor
processes is increased without F-Stack, the DPDK-based network
optimization can slow down the increase of the latencies especially in
the cases of pull latency of Actor and set latency of Learner.
A communication latency reduction can increase net computation time
for action search by Actors and training by Learner.
It can also improve scalability on the number of Actor processes.
In addition, since sending frequency of experiences by Actors is
increased, it is expected that diversity of experiences to be sampled
and trained is enhanced, which also reduces the number of training
epochs required for the convergence.

\subsection{Low-Latency Experience Replay Memory by DPDK}
\begin{figure*}[h]
  \begin{center}
    \includegraphics[height=68mm]{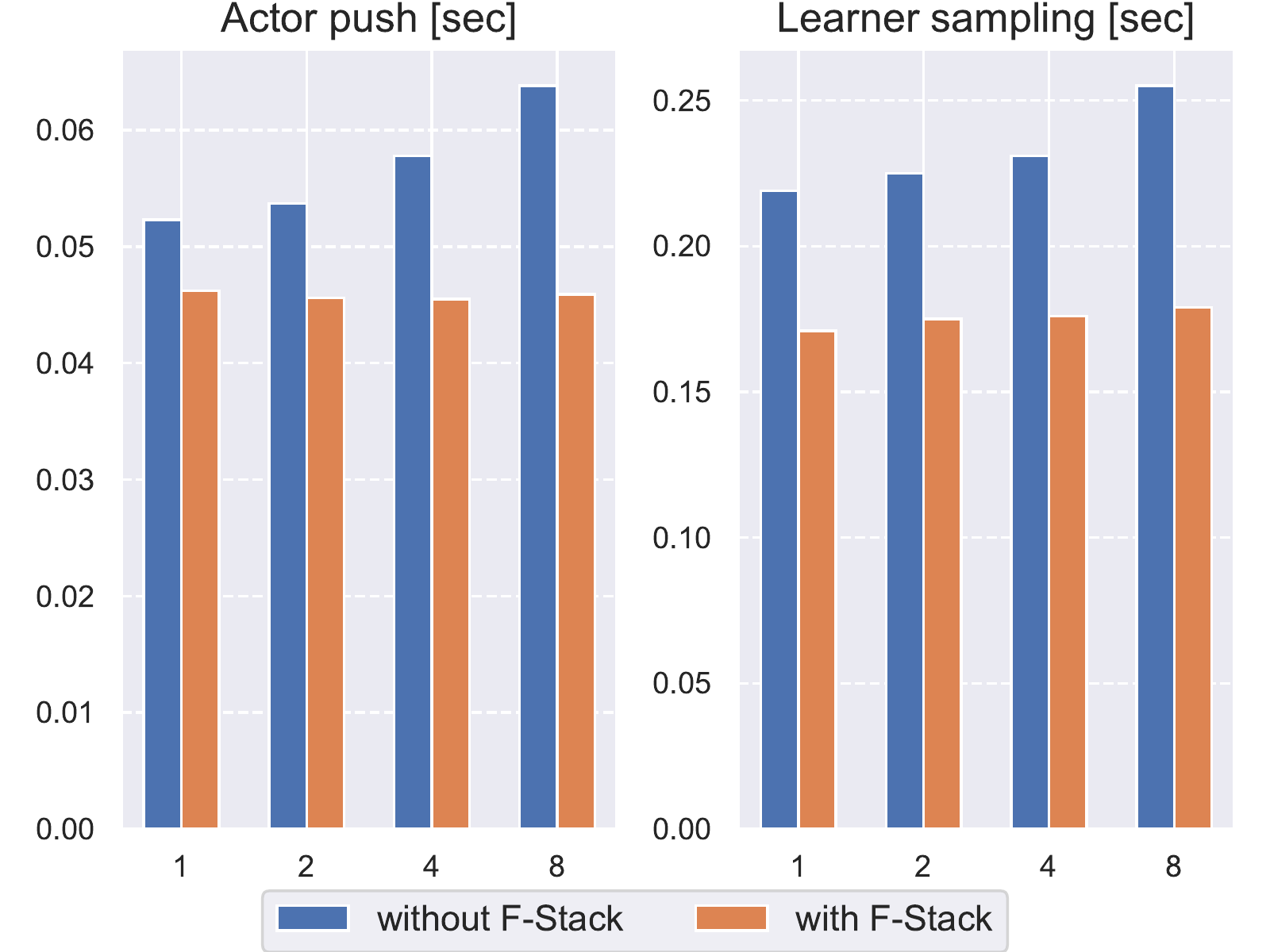}
    \caption{Evaluation result of access latency to experience replay
      memory (second optimized implementation)}
    \label{fig:result2}
  \end{center}
\end{figure*}

As the second optimized implementation, a low-latency experience
replay memory including shared memory implemented with F-Stack is
evaluated in terms of access latencies.
Since an experience replay memory is moved from the Learner node to
the shared memory node in the second optimized implementation, network
access latencies between them are significantly reduced.
The distributed deep reinforcement learning system with and without
F-Stack based network optimization is evaluated in terms of the push
latency of Actor and sampling latency of Learner over network (both of
which have high impact on the latency) when the number of Actor
processes is changed from one to eight.
In Figure \ref{fig:result2}, the left and right graphs show the push
latency of Actor and sampling latency of Learner over network,
respectively.
X-axis is the number of Actor processes, and Y-axis is elapsed time in
second.
By introducing the F-Stack based low-latency experience replay memory,
the push latency of Actor and sampling latency of Learner over network
are reduced by 11.7\%-28.1\% and 21.9\%-29.8\%, respectively.

In the first optimized implementation (i.e., low-latency shared memory
by DPDK), the F-Stack compatible Redis, which is implemented with C
language, is used as a shared memory server, while its experience
replay memory is implemented in Python language.
In the second optimized implementation (i.e., low-latency experience
replay memory including shared memory by DPDK), on the other hand,
since its experience replay memory is newly implemented with C/C++
language, a direct comparison between these two implementations may
not be fair.
Nevertheless, the second optimized implementation can further slow
down the increase of these latencies due to increase of the number of
Actor processes.
Especially, in the second optimized implementation, a dedicated I/O
micro-thread for each Actor is running at the experience replay memory
server (i.e., central node), and it is observed that this further
reduces experience push latency of Actors.
Although micro-thread overheads can be slightly observed when the
number of Actor processes is small, the latency reduction by F-Stack
becomes significant as the number of Actor processes is increased.
That is, the DPDK-based network optimizations become more beneficial
especially with more Actor processes in distributed deep reinforcement
learning systems.
Although only a single CPU core is dedicated to an F-Stack (DPDK)
process in our current implementation, to use more Actor processes,
DPDK throughput should also be increased by introducing a multi-thread
design of using DPDK and RSS (Receive Side Scaling).

\subsection{Discussion}
In our implementation, Actor node, shared memory (experience replay
memory) node, and Learner node are interconnected with a 40GbE
switch. 
In a practical use case, however, Actor processes should be deployed
in edge environment since they are required to tightly interact with
control objects in a field.
We are expecting that our low-latency experience replay memory server
should be located in edge servers and Learner is implemented in
high-performance compute server or cloud.
In our baseline and first optimized implementations, all the
experiences acquired by Actors are transferred to Learner node.
If we assume they are implemented in such edge-cloud architecture, a
large amount of data is transferred between edge and cloud sides via
WAN (Wide Area Network) and network access overheads would be more
significant compared to 40GbE direction connection assumed in our
evaluations, resulting in a lower training efficiency of the
reinforcement learning.
Thus, in addition to reducing network processing overhead by DPDK and
F-Stack, it is required to reduce the communication size between edge
and cloud sides via WAN.
In this context, our second optimized implementation that co-locates
the experience replay memory and shared memory in an edge server is
advantageous since it can reduce the communication size over WAN in
addition to reducing the network processing overhead by DPDK and
F-Stack.
Additional experiments on such edge-cloud environment over WAN are our
future work.


\section{Summary}\label{sec:conc}
To improve performance of deep reinforcement learning such as DQN,
distributed deep reinforcement learning using a cluster of computers
is a promising approach.
In distributed deep reinforcement learning systems, since multiple
nodes in different roles heavily communicate each other, their
communication overheads negatively impact benefits of parallelization
of Actor processes.
In this paper, first, a baseline distributed deep reinforcement
learning system inspired by \cite{apex} was built, and then DPDK-based
low-latency shared memory and experience replay memory servers were
implemented and deployed between Actor and Learner nodes
interconnected with 40GbE.
By introducing the in-network low-latency shared memory server between
Actor and Learner nodes, their network access latencies were reduced
by 32.7\%-58.9\%.
By introducing the in-network experience replay memory server,
experience push latency of Actors was reduced by 11.7\%-28.1\% and
prioritized experience sampling latency over network was reduced by
21.9\%-29.1\%.

Although in this paper, Actor node, shared memory (experience replay
memory) node, and Learner node are located in an ideal 40GbE network,
it would also be a practical configuration where Actor nodes are
distributed at edge environment, the experience replay memory server
is located at an edge server, and high-performance Learner is served
at cloud.
In this case, our second optimized implementation is advantageous
since it can reduce communication size over WAN in addition to
reducing the network processing overhead by DPDK.
As future work, we are currently preparing a long-haul communication
environment using 10km optical fiber cable in order to make additional
evaluations of the proposed in-network shared memory and experience
replay memory servers on such an edge-cloud environment over WAN.
A scalability analysis with more Actor nodes is also our future work.



\end{document}